\documentclass[preprints,article,accept,moreauthors,pdftex]{Definitions/mdpi} 

\firstpage{1} 
\pubvolume{1}
\issuenum{1}
\articlenumber{0}
\pubyear{2022}
\copyrightyear{2021}
\datereceived{} 
\dateaccepted{} 
\datepublished{} 
\hreflink{https://doi.org/} % If needed use \linebreak
\pdfoutput=1

\usepackage{natbib}
\usepackage{url}
\usepackage{fancyhdr}
\usepackage{lastpage}

\Title{Designing a reliable and low-latency LoRaWAN solution for environmental monitoring in factories at major accident risk}

\TitleCitation{Designing a reliable and low-latency LoRaWAN solution for environmental monitoring in factories at major accident risk}

\Author{Dinesh Tamang $^{1}$, Alessandro Pozzebon $^{2}$, Lorenzo Parri $^{1}$, Ada Fort$^{1}$ and Andrea Abrardo$^{1,}$*}

\AuthorNames{Dinesh Tamang, Alessandro Pozzebon, Lorenzo Parri, Ada Fort and Andrea Abrardo}

\AuthorCitation{Tamang, D.; Pozzebon, A.; Parri, L.; Fort A.;Abrardo, A.}
\address{%
$^{1}$ \quad Department of Information Engineering and Mathematical Science, University of Siena, Italy; dinesh.tamang@student.unisi.it; parri@diism.unisi.it; abrardo@diism.unisi.it, ada@diism.unisi.it\\
$^{2}$ \quad Department of Information Engineering, University of Padova,  35131 Padova, Italy; alessandro.pozzebon@unipd.it
}

\corres{Correspondence: abrardo@diism.unisi.it}

\abstract{In this article, we propose a reliable and low-latency Long Range Wide Area Network (LoRaWAN) solution for environmental monitoring in factories at major accident risk (FMAR). In particular, a low power wearable device for sensing the toxic inflammable gases inside an industrial plant is designed with the purpose of avoiding peculiar risks and unwanted accidents to occur. Moreover, the detected data have to be urgently and reliably delivered to remote server to trigger preventive immediate actions so as to improve the machine operation. In these settings, LoRaWAN has been identified as the most proper communications technology to the needs owing to the availability of off the shelf devices and software. Hence, we assess the technological limits of LoRaWAN in terms of latency and reliability and we propose a fully LoRaWAN compliant solution to overcome these limits. The proposed solution envisages coordinated end device (ED) transmissions through the use of Downlink Control Packets (DCPs). Experimental results validate the proposed method in terms of service requirements for the considered FMAR scenario.}

\keyword{LoRaWAN; Reliability; Downlink; Safety; IoT} 

\begin{document}
%%%%%%%%%%%%%%%%%%%%%%%%%%%%%%%%%%%%%%%%%%

\section{Introduction}
%\subsection{The application scenario}
\label{appliation_scenario}

Safety can be defined as the state of being free from unacceptable risks which can potentially cause damages to humans, environment or properties \cite{IEC_TR}. Public safety is very important in industrial environments especially in  high-risk fields such as oil and gas, chemical industry or nuclear reactor plants where any event or accident can lead to catastrophic consequences both for humans, i.e., the workers and/or those who are living in the surrounding areas, and for the environment. These kinds of scenarios are typically categorized as \emph{Factories at Major Accident Risk} (FMAR). A mapping of the most frequent major accidents in this context has been produced in Italy in 2013 by the ISPRA (Istituto superiore per la protezione e la ricerca ambientale - High Institute for Environmental Protection and Research) \cite{ISPRA} and they are mainly due to release of dangerous substances (liquid or gas), fires, and explosions. The possible causes are several, such as machinery or piping break up, watertight loss, tanks overfill, mistakes during the handling procedures, submission of material on already filled cans, blending of incompatible materials, valves break up, accidental falls or vehicle collisions, and, more generally, spillover of cisterns.
    
    Activities of ensuring compliance with procedures to prevent major accidents are traditionally in charge of humans control, and, as such, they are naturally error prone. On the other hand, with the rapid development of the numerous innovations in manufacturing and in information and wireless communications technologies, we have assisted in the last years to the birth of the new Industry 4.0 era \cite{Chen2017}. In this scenario, the Internet of Things (IoT) is a hot topic nowadays either from a research point of view and from an application perspective as well \cite{Palattella2016}. Needless to say, IoT may provide great support to safety in the considered FMAR scenario in which workers and machines operate in a shared environment and a great amount of heterogeneous information can be collected by sensors and automatically delivered to a Central Controller (CC) allowing a real time control of the whole activity chain (and, as a consequence, of the associated risks). More specifically, the most important sensing technologies in the considered application scenario are represented by sensors for detecting gases with the purpose of avoiding fires and explosions involving flammable gas leakages as well as for controlling the level of oxygen in the atmosphere. For this purpose, we refer to the wearable device embedding sensing and communications capabilities described in Section \ref{SC}, which is under development within a collaboration between the University of Siena and INAIL (Istituto nazionale per l'assicurazione contro gli infortuni sul lavoro - Italian National Institute for Insurance against Accidents at Work). Regarding in particular the communication aspects, the considered IoT scenario is characterized by low power devices transmitting infrequent short bursts of data over a low power wide area network. Indeed, the devices cannot have any external power supply (i.e., they run on batteries and they are installed in areas where frequent batteries substitution cannot be always guaranteed). Accordingly, they are expected to be very power efficient. Moreover, most of the detected data are not critical and, as such, can be referred to as \emph{Regular Packets} (RPs). However, in some cases, i.e., when the concentration of gases crosses a pre-defined threshold, the detected data have to be urgently and reliably delivered and, accordingly, they are referred to as \emph{Urgent Packets} (UPs). In this setting, one of the most promising communications technologies is represented by the Long Range (LoRa) one, together with the associated LoRa Wide Area Network (LoRaWAN) protocol.

When designing a LoRa-based network infrastructure, the adoption of LoRaWAN  provides several advantages with respect to other customized Media Access Control (MAC) protocols. First of all, there is a large availability of open source LoRaWAN servers which can be easily installed and employed to rapidly set-up LoRaWAN networks. As a matter of fact, most of the Gateways (GWs) currently available on the market support the LoRaWAN protocol. In terms of range and coverage, this technology provides a far longer range than Wireless Fidelity (WiFi) or Bluetooth connections \cite{Mekki2019}, applicable for indoor as well as outdoor scenarios, especially in remote areas where the cellular networks have poor connections. Moreover, when setting up dense network infrastructures, upper layer protocols like MAC ones are developed in LoRaWAN with the aim of managing the whole network infrastructure as well as the coexistence of large quantity of end devices (EDs). Moreover, LoRaWAN provides several built-in features that can be very important for the scenario at hand such as: (\emph{i}) security in the transmission by means of Advanced Encryption Standard (AES-128) end-to-end data encryption; (\emph{ii}) Possibility of boosting the capacity through the use of multiple channels; (\emph{iii}) Adaptive Data Rate (ADR) and power consumption by controlling the Spreading Factor (SF), the Bandwidth (BW) and the Coding Rate (CR). 
 
Finally, LoRaWAN provides a number of different message types which allow to set up unconfirmed (UNCONF) or confirmed (CONF) data transmissions (by means of Acknowledgement (ACK)) and a downlink (DL) channel to send information back to ED as well as the Over-The-Air Authentication (OTAA), a procedure that simplifies the association of an ED to the network by means of Join request messages. 

The rest of the paper is structured as follows. In Section \ref{section:related_works}, the state of the art of LoRa solutions for reliable communication and related works are presented. In Section \ref{SC}, we introduce the sensing and communication parts describing the main sensor characteristics, power consumption analysis and server architecture. Then, the proposed approach is described in Section \ref{section:proposed-approach}. The main experimental results and discussions are presented in Section \ref{section:results and discussion}. Finally, concluding remarks are given in Section \ref{Conclusion}.

%%%%%%%%%%%%%%%%%%%%%%%%%%%%%%%%%%%%%%%%%%
%\section{Related Works}

\section{State of the art}
\label{section:related_works}
Various works have investigated the reliability of LoRa networks. In \cite{collision-free-Hairahem}, the authors investigate the problem of collision and propose two distinct mechanisms for collision free transmission, namely TDMA
(Time-Division Multiple Access)-based and FDMA (Frequency Division Multiple Access)-based with an ultimate aim of increasing the reliability of the service. The first mechanism allows all clusters to transmit in sequence where up to six EDs belonging to the same cluster can transmit using different SFs in parallel whereas the latter allows all clusters to transmit in parallel, each cluster on its own frequency. However, within each cluster, all EDs transmit in sequence. The simulation results provide better performances than standard LoRaWAN in terms of Packet Delivery Rate (PDR) even if the number of EDs is high. Similarly in \cite{B.Reynders}, a two-step lightweight scheduling is proposed to divide nodes into groups where similar transmission powers are used in each group to reduce the capture effect. The nodes are guided by the GW coarse-grained scheduling to use different SFs to enable simultaneous transmissions through the use of beacon signals at every pre-defined interval, thus reducing packet collisions. The validity of the proposed scheme is assessed using NS-2 simulations showing better performance than legacy LoRaWAN in terms of packet error ratio, throughput, and energy efficiency. However, inter-SF transmission is still a problem due to the loss of the orthogonality between the two signals \cite{inter-SF interference} \cite{imperfect orthogonality}. Since the ALOHA mechanism of the LoRaWAN drastically decreases the performance because of the non-negligible on-air collision probabilities, some authors in literature have proposed the synchronization of LoRa networks by assigning slots to each node using fine-grained scheduling \cite{Abdelfadeel}. 

In order to overcome the problems of classical ALOHA in LoRaWAN, Zorbas et al. \cite{Zorbas} propose a time-slotted mechanism where data are buffered locally and transmitted whenever a GW is available by avoiding bursts of collisions. Similarly, in \cite{TS-LoRa} a Time Slotted (TS)-LoRa that allows the nodes to organize autonomously and determine their slot positions in a frame is proposed. This is achieved by sharing an easy-to-compute hash algorithm between the network server and the nodes able to map the nodes’ addresses that are assigned during the join phase into unique slot numbers. Moreover, this mechanism ensures backward compatibility with legacy LoRaWAN nodes and liberates TS-LoRa from the huge schedule dissemination overhead. The last slot in each frame is used for sending synchronization ACK responsible for handling time synchronization and ACKs. The considered TS-LoRa achieves a very high packet delivery ratio for all the tested SFs.

The availability of CONF messages is one of the important features in the LoRaWAN networks that is not available in some of its competing low power technologies like Sigfox, Bluetooth Low Energy (BLE) etc. This feature can be used in those scenarios where data reliability is concerned. However, very few works in literature have investigated the CONF traffic, its effects and, more in general, the DL viability. Marais et al. \cite{Marais} provides an analysis of use cases requiring CONF traffic and concludes that CONF traffic is viable in small networks, especially when data transfer is infrequent. Additionally, aspects likes duty-cycle regulations, SF12 for RX window 2, maximum re-transmission numbers and ACK\_TIMEOUT transmission back-off interval negatively impact the viability of the CONF traffic. Similarly, Capuzzo et al. \cite{Capuzzo} conclude that the performance of a single LoRaWAN cell can significantly degrade when the fraction of nodes that require CONF traffic grows excessively. Moreover, they also suggest that it is necessary to carefully choose the maximum number of transmission attempts for CONF packets, based on the node density and traffic load to get the best performances. In addition, various works in the literature \cite{Farhad,Varsier,Markkula,Pop} investigate the applicability as well as criticism of DL in LoRaWAN networks and its possible negative impact on performances when not well implemented.
To sum up, LoRaWAN technology has limitations that need to be carefully considered for its use in the considered FMAR scenario. In particular, one of the most critical issues is related to the use of the CONF mode to provide link reliability. Indeed, as discussed in \cite{Adelantado}, the use of ACK in DL can significantly drain the network capacity since GWs must be compliant with duty-cycle regulations. This problem is also studied in detail in \cite{Magrin} where the authors propose a solution called \emph{sub-band swapping} where a first receiving window is opened on the dedicated downlink (DL) channel and a second one on the uplink (UL) channel to alleviate the duty cycle's bottleneck. Another line of investigation deals with the use of ad-hoc control schemes which deviate from the standard LoRaWAN ACK mechanism \cite{Hasegawa,Centenaro}. 

In this paper, we propose a reliable and low-latency communications solution which is suitable for the considered FMAR scenario and that is fully LoRaWAN compliant. On the other hand, to the best of our knowledge all the previous works in the literature addressing the problem of reliability in LoRa deviate from LoRaWAN standard and, as such, would require a brand-new firmware update at both nodes' and GW level. Conversely, the aim of the solution proposed in this work is to integrate the standard LoRaWAN configurations: this would make the implementation of this system almost straightforward since only minor modifications on the server side of the network infrastructure are required. As a matter of fact, the proposed solution may exploit already existing LoRaWAN networks and thus may be installed without any need for major infrastructural integration. Accordingly, this work proposes a fully operating experimental setup and the viability of the proposed scheme is demonstrated by means of a fully operating LoRaWAN network infrastructure whereas in all the referenced works the results could be achieved only by means of simulations. 

\section{The Integrated Sensing and Communication Platform}\label{SC}
\subsection{Sensor Node Architecture}
The sensor node was fully custom designed to fulfill the requirements of the application scenario and to comply with the constrains in terms of physical dimensions, measurement accuracy and energy consumption. It must be wearable so its dimensions and its weight must be reduced as much as possible to allow the attachment of the device to the belt or the helmet of an operator. As for energy supply, we refer to a standard 18650 type lithium battery. To this regard, the power consumption has been optimized to guarantee a battery lifetime up to one working week (40 hours). Finally, the long transmission range, which is fundamental since the system is expected to be deployed also in large industrial sites, is ensured by the LoRaWAN connectivity.

\begin{figure}[ht]
	\centering
	\includegraphics[width=1.0\columnwidth]{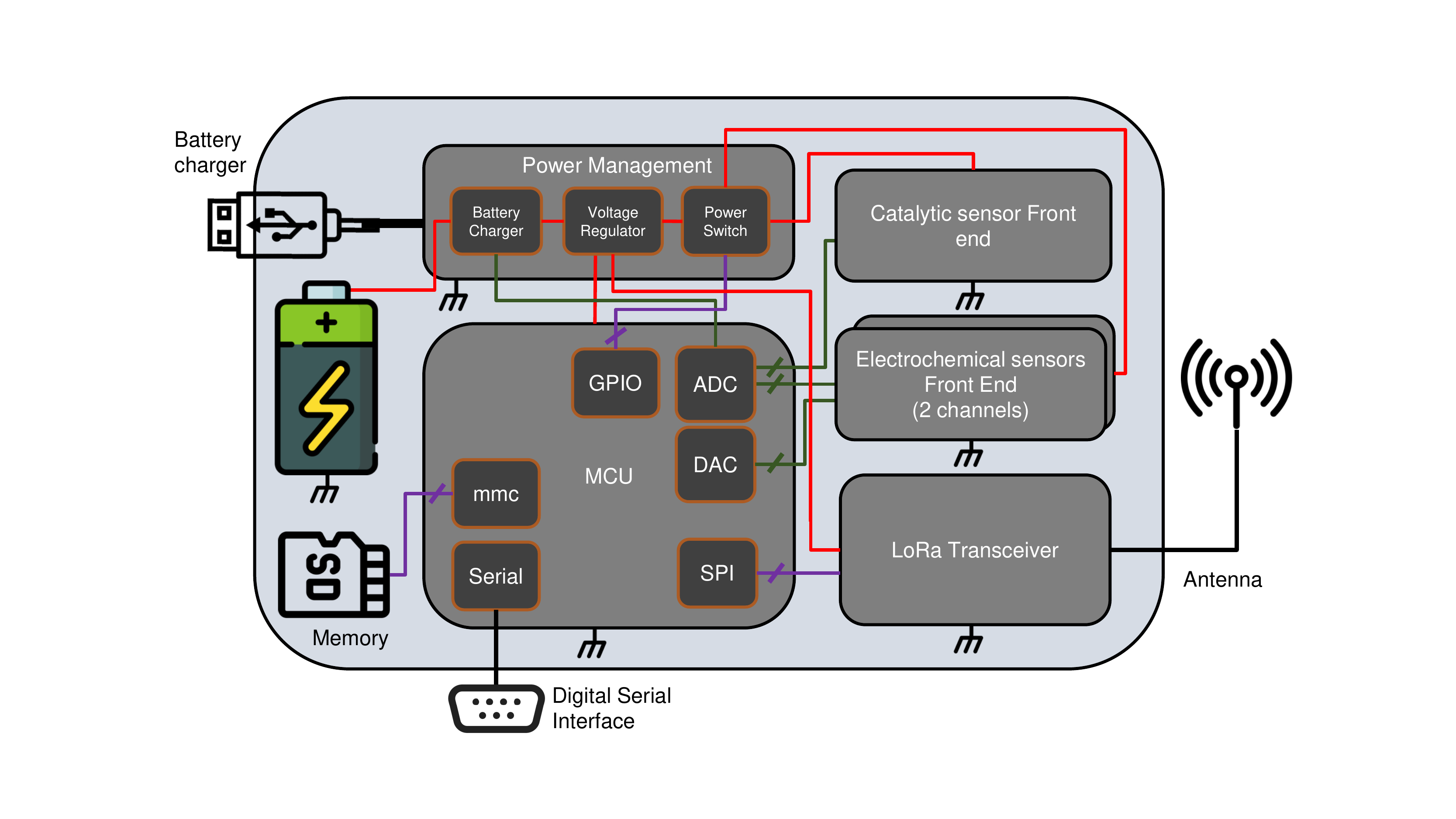}
	\caption{Sensor node architecture.}
	\label{fig:sensornode-architecture}
\end{figure}

\begin{figure}[ht]
	\centering
	\includegraphics[width=0.9\columnwidth]{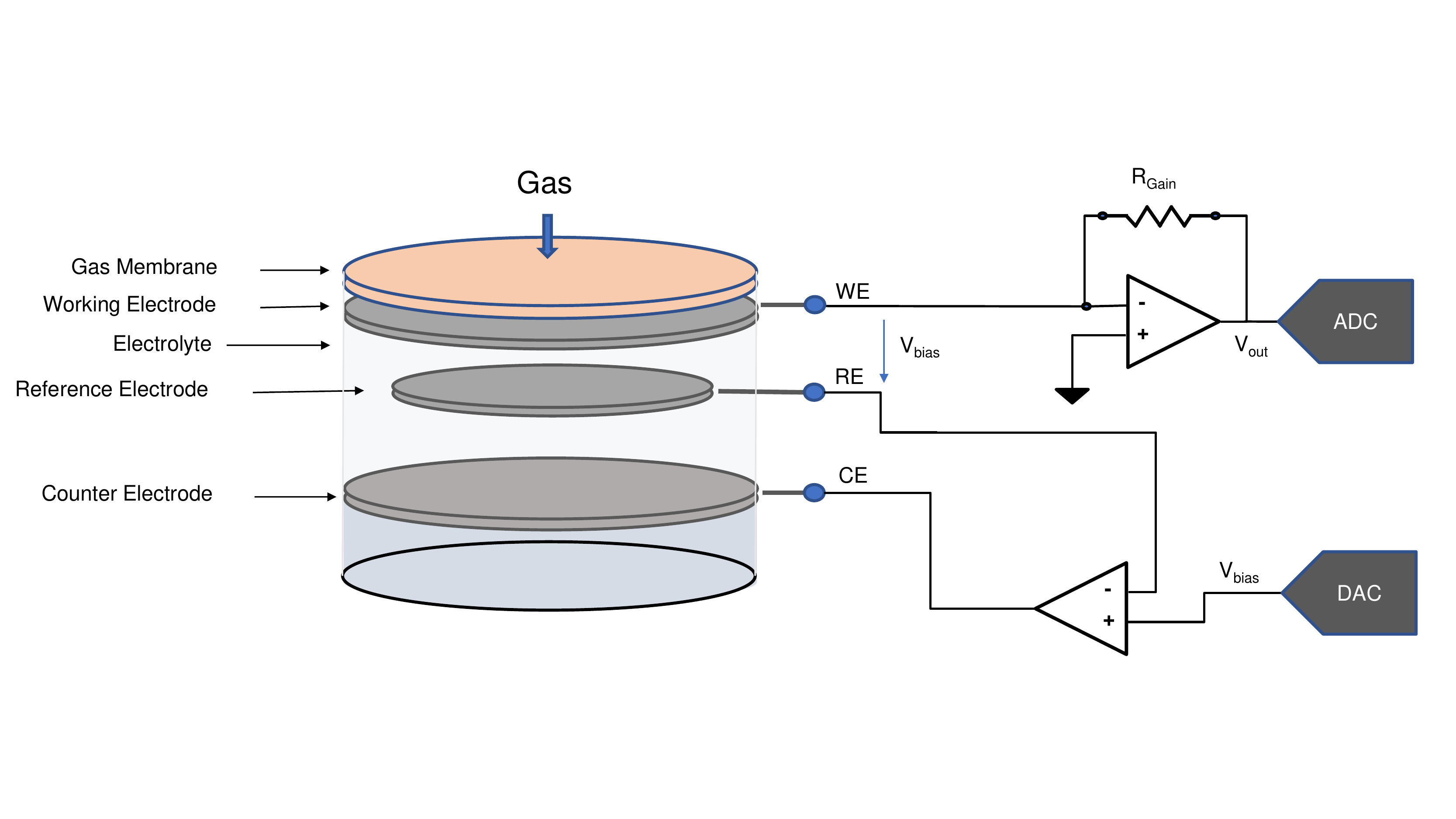}
	\caption{Electrochemical Sensors front end circuit.}
	\label{fig:frontend-electrochem}
\end{figure}

The sensor node architecture is reported in Figure \ref{fig:sensornode-architecture}. It encompasses 3 gas sensors, two electrochemical amperometric gas sensors and a catalytic gas sensor. The electrochemical sensors front end circuit is based on a standard potentiostatic circuit with an I-V converter as the one presented in \cite{CarbonMonoxTrig}, allowing to host 4 electrodes sensors with an adjustable biasing value as shown in Figure \ref{fig:frontend-electrochem}. The sensors used in this application are the CO-A4 for the measurement of carbon monoxide concentration (CO) and the O2-A1 for the measurement of oxygen concentration (O$_2$), both from Alphasense.

The CO-A4 sensor has been calibrated to detect dangerous concentrations of CO (up to 500 ppm with a resolution of $\pm2\,ppm$), while the O2-A1 to detect oxygen-deficient atmospheres (range 15\%-21\% with a resolution of $\pm0.5\%$). The catalytic sensor is the CH-A3 from Alphasense, for the detection of potentially explosive atmospheres where the concentration of methane, propane or butane is above the Lower Explosive Limit (LEL). The LEL is the minimum concentration of an explosive gas in air that makes possible an explosion in presence of a trigger. The catalytic sensor is composed of a pellet of catalyst loaded ceramic whose electrical resistance changes in presence of combustible gases. The ceramic pellet needs to be heated by flowing a current, to catalyze a chemical reaction that involves the gas on its surface. The pellet resistance may significantly vary with the temperature, for this reason the sensor contains two identical sensing elements: one of the two is not exposed to the gas and it is used as a reference element.
\begin{figure}[ht]
	\centering
	\includegraphics[width=0.9\columnwidth]{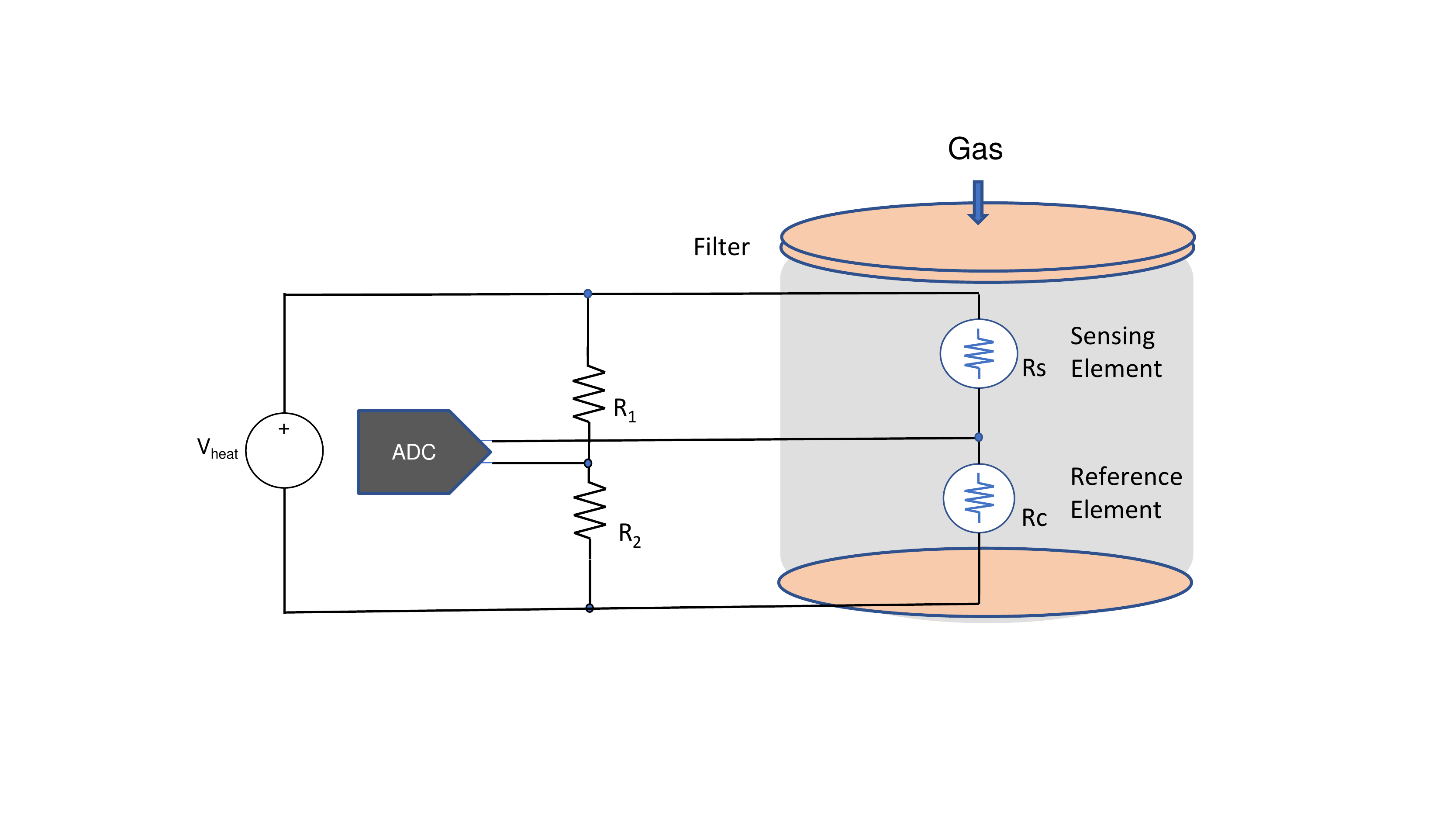}
	\caption{Catalytic Sensor front end circuit.}
	\label{fig:frontend-pellistor}
\end{figure}

By realizing a Wheatstone bridge, as shown in Figure \ref{fig:frontend-pellistor}, composed by the two elements (the one exposed to the gas and the reference) in a branch and two reference resistors in the other, it is possible to compensate the temperature effect and the gas concentration is obtained by measuring the output voltage of the bridge. The sensor exhibits a different sensitivity to butane, propane and methane: the sensitivity of the sensor toward methane is the lowest one (butane and propane sensitivity is 150\% of methane sensitivity from producer specifications). The LEL for methane is 5\% while for butane is 1.8\% and for propane is 2.1\% in volume. By calibrating the sensor only in methane, it is possible to set a safe voltage level to detect the risk of explosion. For example, a sensor with a sensitivity of 2\%/V in methane, exhibits a sensitivity toward propane and butane of 3\% /V. Considering the LEL values for the target gases, we obtain that the LEL of methane corresponds to 2.5 V (5\%), the one of butane (1.8\%) to 0.6 V whereas the LEL of propane (2.1\%) is 0.7 V. Setting an alarm threshold of 0.5 V allows to obtain a good safety margin for all these gases. 

The digital part of the node is based on a low power ARM microcontroller from STMicroelectronics (STM32LQT5), that embeds 12 bits Analog to Digital Converters (ADCs) and Digital to Analog Converters (DACs) used to acquire signals from the sensors front end and to provide the correct biasing of electrochemical sensors. The microcontroller can save and read data from an Secure Digital (SD) card memory storage for logging purposes and to load sensors calibration parameters. The node sends data trough the LoRa radio channel exploiting an RFM95 transceiver by HopeRF, interfaced through an SPI bus.

\begin{figure}[ht]
	\centering
	\includegraphics[width=0.4\columnwidth]{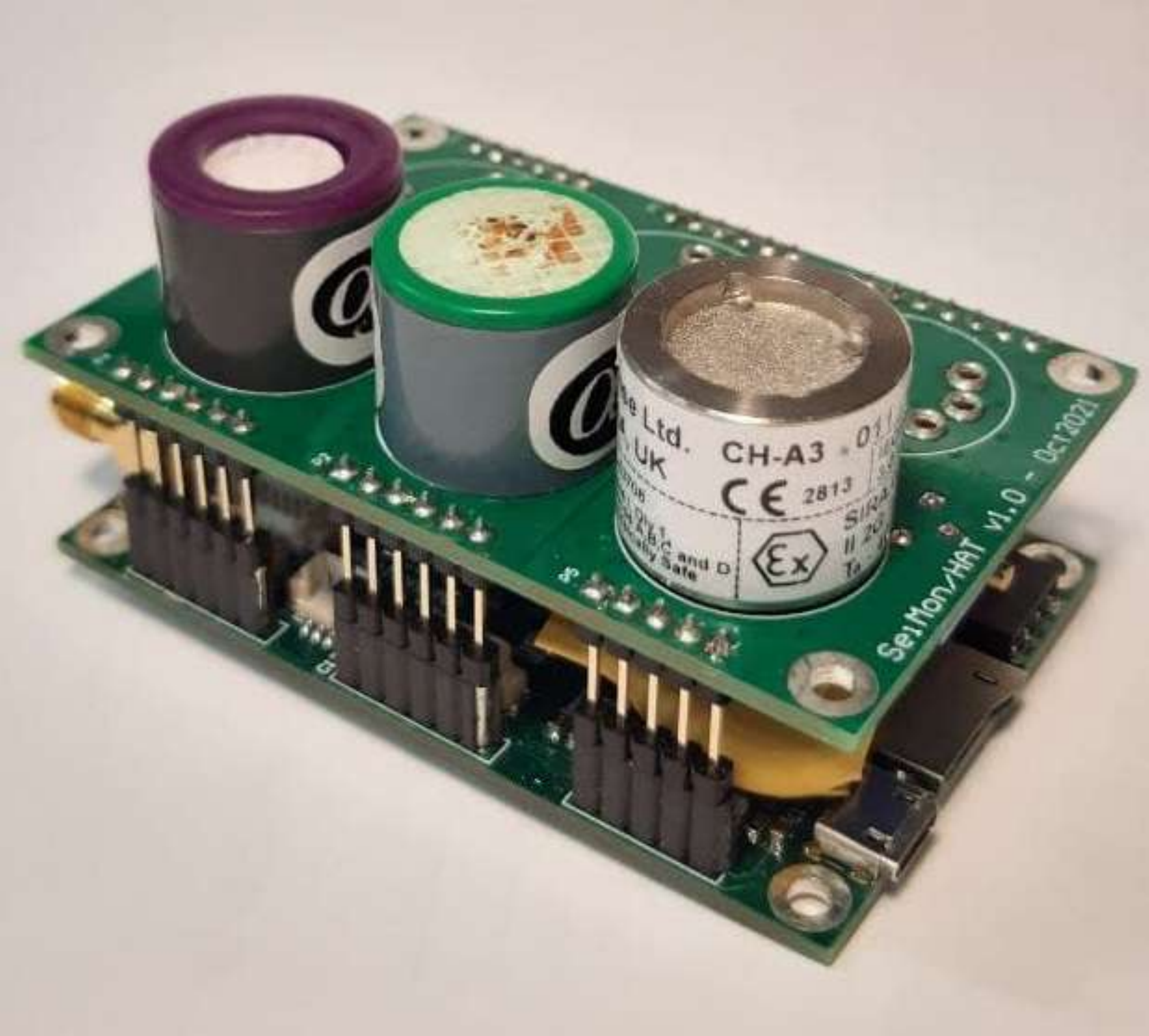}
	\caption{Sensoring device without enclosure }
	\label{fig:Fonotonodo-Elett}
\end{figure}

\begin{figure}[ht]
	\centering
	\includegraphics[width=0.4\columnwidth]{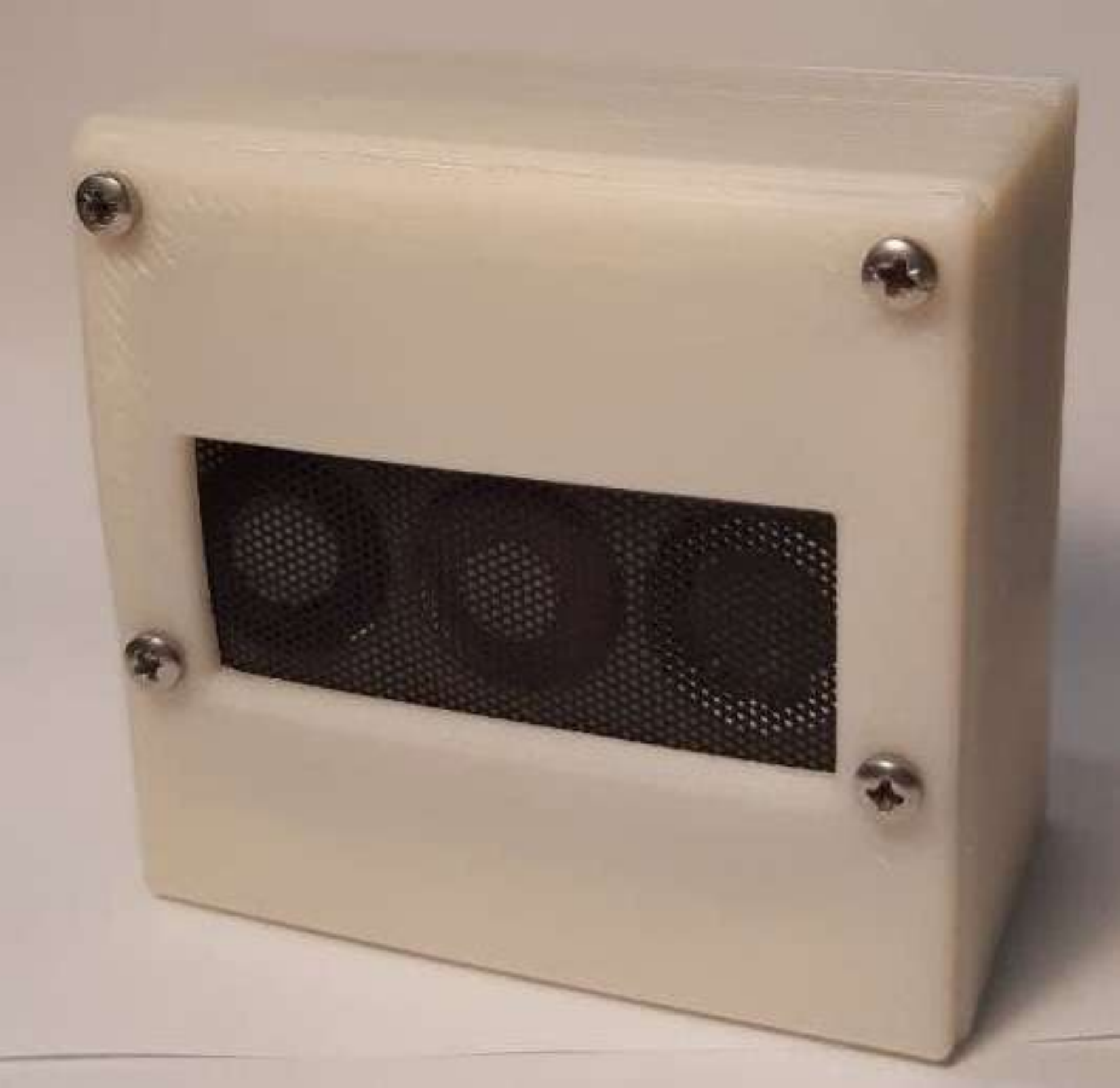}
	\caption{Sensoring device with enclosure }
	\label{fig:Fonotonodo-Enc}
\end{figure}

\begin{figure}[ht]
	\centering
	\includegraphics[width=0.4\columnwidth]{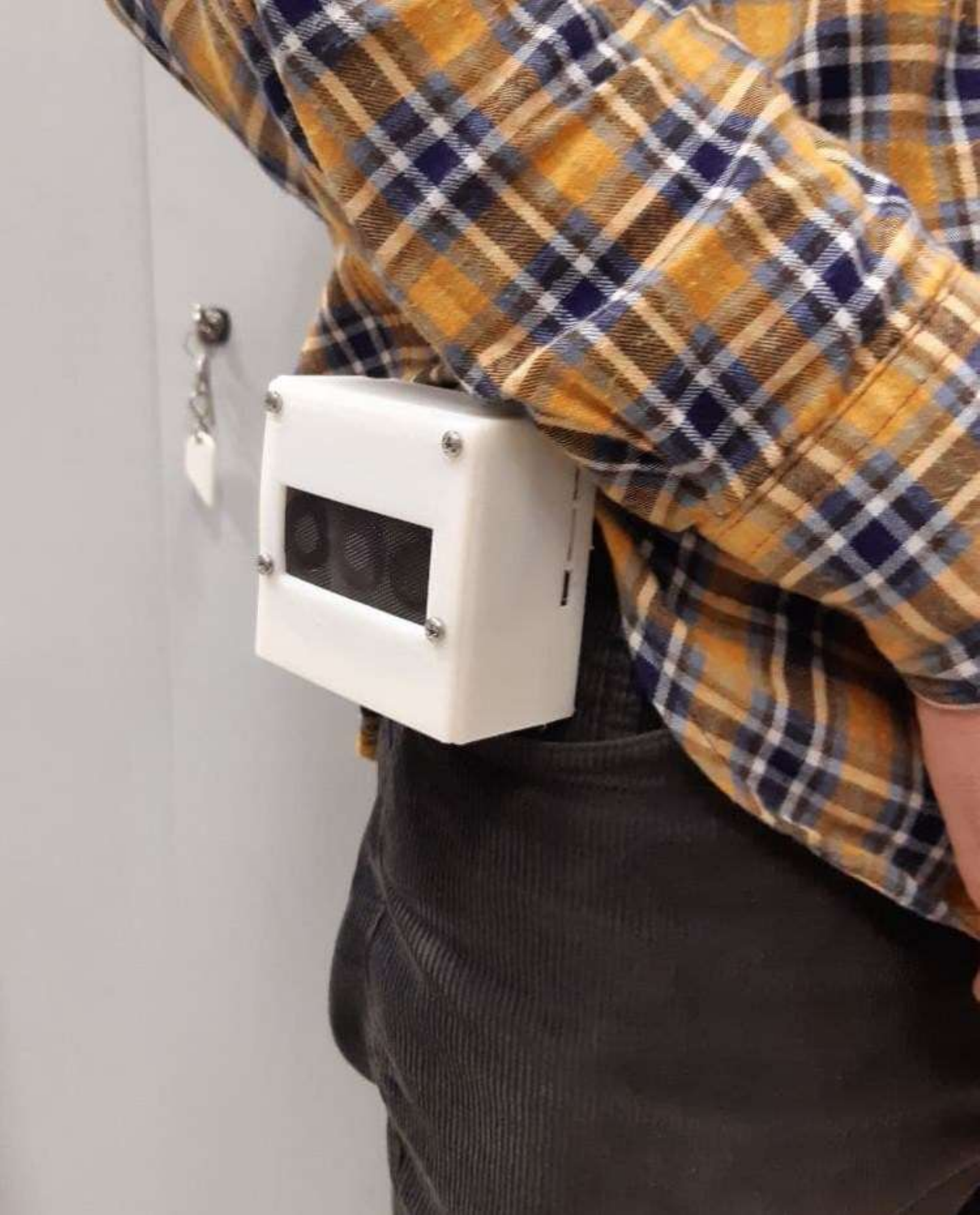}
	\caption{Sensoring device fitted to the belt }
	\label{fig:Fonotonodo-Cint}
\end{figure}

\subsection{LoRaWAN communication platform: LoRa node, GW and server}
 In the following we provide a brief outline on the overall technology and the network architecture considered in this work.\\
 LoRaWAN protocol is based on the LoRa transmission technology, a proprietary modulation patented by Semtech. LoRa operates in the unlicensed Sub-GHz (below 1 GHz) Industrial, Scientific and Medical (ISM) bands, with three operating frequencies: 433 MHz, 868 MHz and 915 MHz. It exploits the Chirp Spread Spectrum (CSS) modulation technique which allows to achieve extremely high receiver sensitivity values (up to -146 dBm): in these conditions, very long transmission ranges are obtained (up to some kms in urban areas and some tens of kms in rural areas) with limited power consumption. LoRa is then the ideal candidate for a plethora of wide area IoT applications, with a large number of connected devices. LoRaWAN networks adopt a star-of-stars topology, which enables multiple GWs to receive packets from a large quantity of EDs: GWs are in charge to transfer the packets to a central network server which manages the network aspects related to security, scalability and reliability. EDs are divided in 3 Classes (Class A, Class B and Class C) which differ for their ability to receive DL packets from the GW. Class A devices are the simplest and less power hungry ones and, as such, they are by far the most common ones. However, all the three ED typologies are bi-directional in operation. Every ED must be registered with a network before performing communication. These activation processes are of two types:  a) OTAA; the most secure and recommended for EDs and b) Activation by Personalization (ABP); less secure and requires hardcoding the device address as well as the security keys in the device. Moreover, one of the important aspects of LoRaWAN is the use of frequency plan and its duty cycle regulations. More specifically, duty cycle indicates the fraction of time a resource is busy. As an example, when a ED transmits on a channel for 3 time units every 10 time units, the device has a duty cycle of 30\%. As for the transmission frequencies, they are specified in the LoRaWAN regional parameters document \cite{lorawan-regional-parameter}. Note that the words frequency and channel are used interchangeably throughout the paper. The duty cycle policy is often regulated by the government and it applies to an entire sub-band. This means that if a user transmits in one of the channels of a given sub-band, it cannot use any of the frequencies of the same sub-band for a time interval regulated by the duty cycle policy. Specifically, the duty cycle values for different sub-bands are regulated by the ETSI EN300.220 standard and are reported below \cite{ETSI300.220};.  
  \begin{itemize}
 \item g (863.0 – 868.0 MHz): 1\%
 \item g1 (868.0 – 868.6 MHz): 1\%
\item g2 (868.7 – 869.2 MHz): 0.1\%
\item g3 (869.4 – 869.65 MHz): 10\%
\item g4 (869.7 – 870.0 MHz): 1\%
 \end{itemize}
 
 The above regulations apply to both EDs and GWs. 
In the followings, we briefly describe the main components of overall system.

\subsubsection{RFM95 Radio Transceiver}
RFM95 LoRa module is a radio transceiver manufactured by HopeRF \cite{rfm}. It has a receiver sensitivity of -148 dBm and power amplifier of +20 dBm. Consequently, it has a maximum link budget of 168 dBm. It requires 3.3 V of voltage supply and draws a minimum RX current of 10.3 mA. The choice of this transceiver is due to its low cost and its low power consumption with a very good receiver sensitivity, suitable for the proposed application scnario. %allowing to make feasible to use in our scenario.
\subsubsection{LPS8 GW}
LPS8 is an open source LoRaWAN GW \citep{lps8} which acts as a bridge between the ED and the network infrastructure. It has a backhaul Internet connectivity that connects it to the remote network server. The LPS8 uses a Semtech packet forwarder, a software responsible for forwarding packets to the server and includes a SX1308 LoRa concentrator. It allows users to send data and reach extremely long ranges at low data-rates providing 10 parallel demodulation paths. The receiver has a sensitivity of up to -140 dBm with SX1257 Tx/Rx front-end.

\subsubsection{Chirpstack LoRaWAN Server}
We consider ChirpStack open-source LoRaWAN Network Server stack \cite{chirpStack} for the server side. Chirpstack provides open-source components to form the network infrastructure. Any instance of each component can be installed locally or in a cloud platform to construct the overall network infrastructure. Moreover, this infrastructure provides a user-friendly web-interface for device management and Application Programming Interfaces (APIs) for integration.

\begin{figure}[ht]
	\centering
		 \includegraphics[width=0.8\columnwidth]{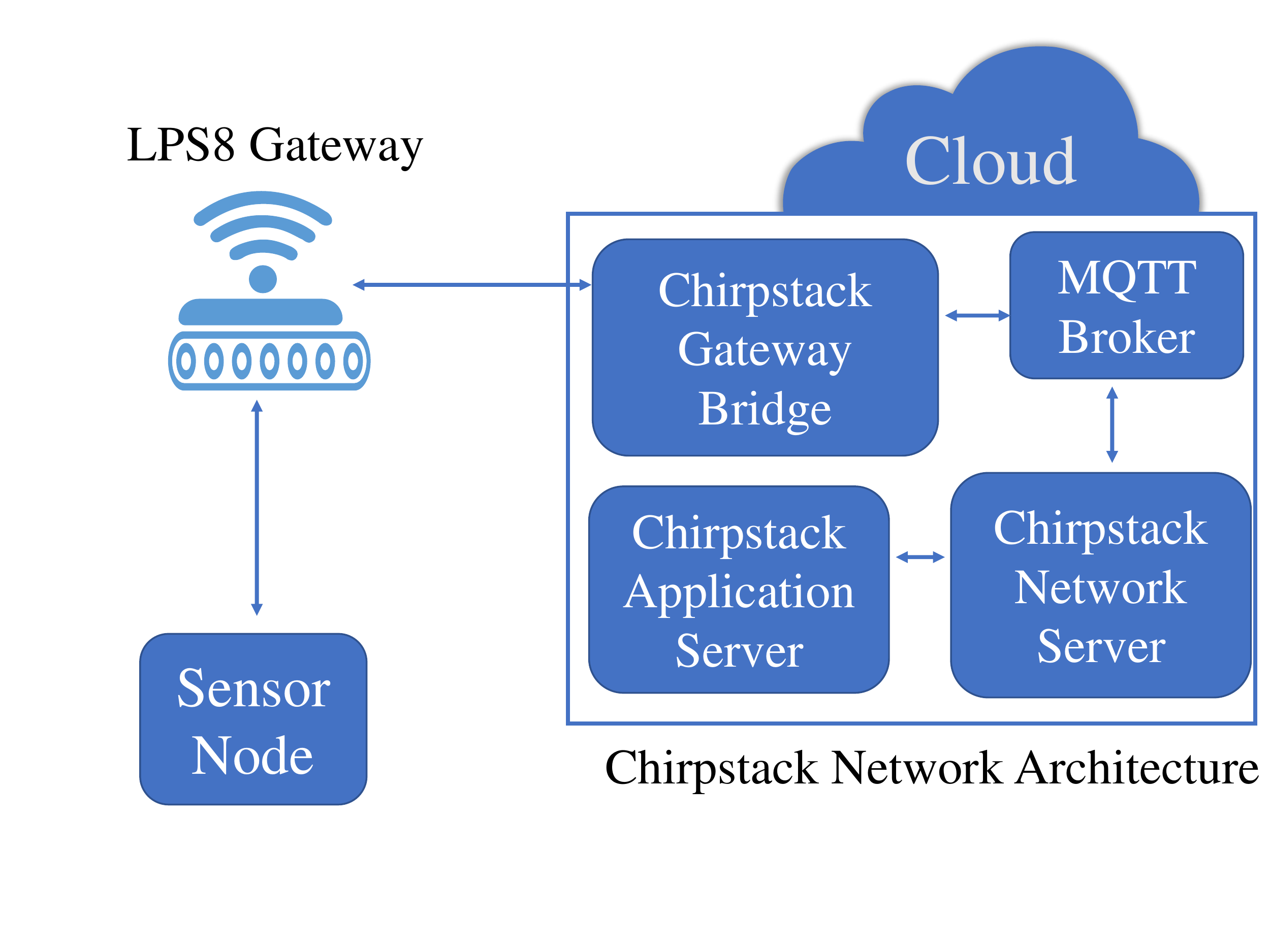}
		\caption{LoRaWAN Infrastructure together with Sensor Node shown in Figure \ref{fig:sensornode-architecture}.}
		\label{fig:lorawan-architecture}
\end{figure}

It is also important to highlight that by default Chirpstack uses Message Queue Telemetry Transport (MQTT) protocol for publishing and receiving application payloads. MQTT is used by ChirpStack GW Bridge, ChirpStack Network Server, and ChirpStack Application Server. Figure \ref{fig:lorawan-architecture} provides the overall network architecture including the sensor node/ED described in the previous subsection.

\subsection{Communication Service requirements}
\label{subsection:service requirement}
As discussed in detail in Section \ref{appliation_scenario}, in order to avoid fires and explosions in the presence of faults and gas leakages, it is necessary to promptly communicate to the CC any anomalous concentration of gases. In the considered system, this kind of data is referred to as UPs. Accordingly, UPs are characterized by stringent requirements in terms of reliability and latency. 

More specifically, in the scenario at hand we have identified as minimum requirements a packet loss rate (PLR) and end-to-end latency (\textit{l}) of $0.1\%$ and  $500 ms$, respectively. Moreover, basing on the required resolution of the data collected from the sensors, the UPs payload is set to 24 bytes. Basing on these requirements, we have limited our LoRaWAN system to use only SF7-10, since with SF11 and SF12 the time on air exceeds 500 ms \cite{air-time-cal}. Finally, packets retransmissions are not allowed for UP packets which, as such, can be transmitted in UNCONF mode only.

\section{The Proposed Communication Strategy}
\label{section:proposed-approach}
In the following, we will focus on the approach proposed in this paper to provide the required reliability. To this aim, we exploit the DL communication scheme provided by standard LoRaWAN: DL packets are sent by a Network Server to only one ED through one or more GWs. To elaborate, in the proposed system a remote central LoRaWAN server shown in Figure \ref{fig:lorawan-architecture} is capable of performing various tasks such as reception of data from the EDs forwarded by GWs, exploitation of the collected data with further processing, and more importantly scheduling of DL messages to the EDs for enabling coordinated transmissions. We describe each aspect in detail in the following subsections.

\subsection{Clustering of EDs}
\label{subsec:positioning system}
In our system, the central server not only collects the sensor data from the EDs, but it is also responsible of forming clusters of users in close proximity. This task is achieved assuming that the system is equipped with a localization system where the server is aware of users' locations. More specifically, the creation of clusters of users is performed with the aim of coordinating the transmissions of close users to avoid possible packets collisions. Indeed, it is highly probable that a critical event such as the presence of gas is jointly detected by all the EDs of the cluster. 

\begin{figure}[ht]
	\centering
		 \includegraphics[width=0.8\columnwidth]{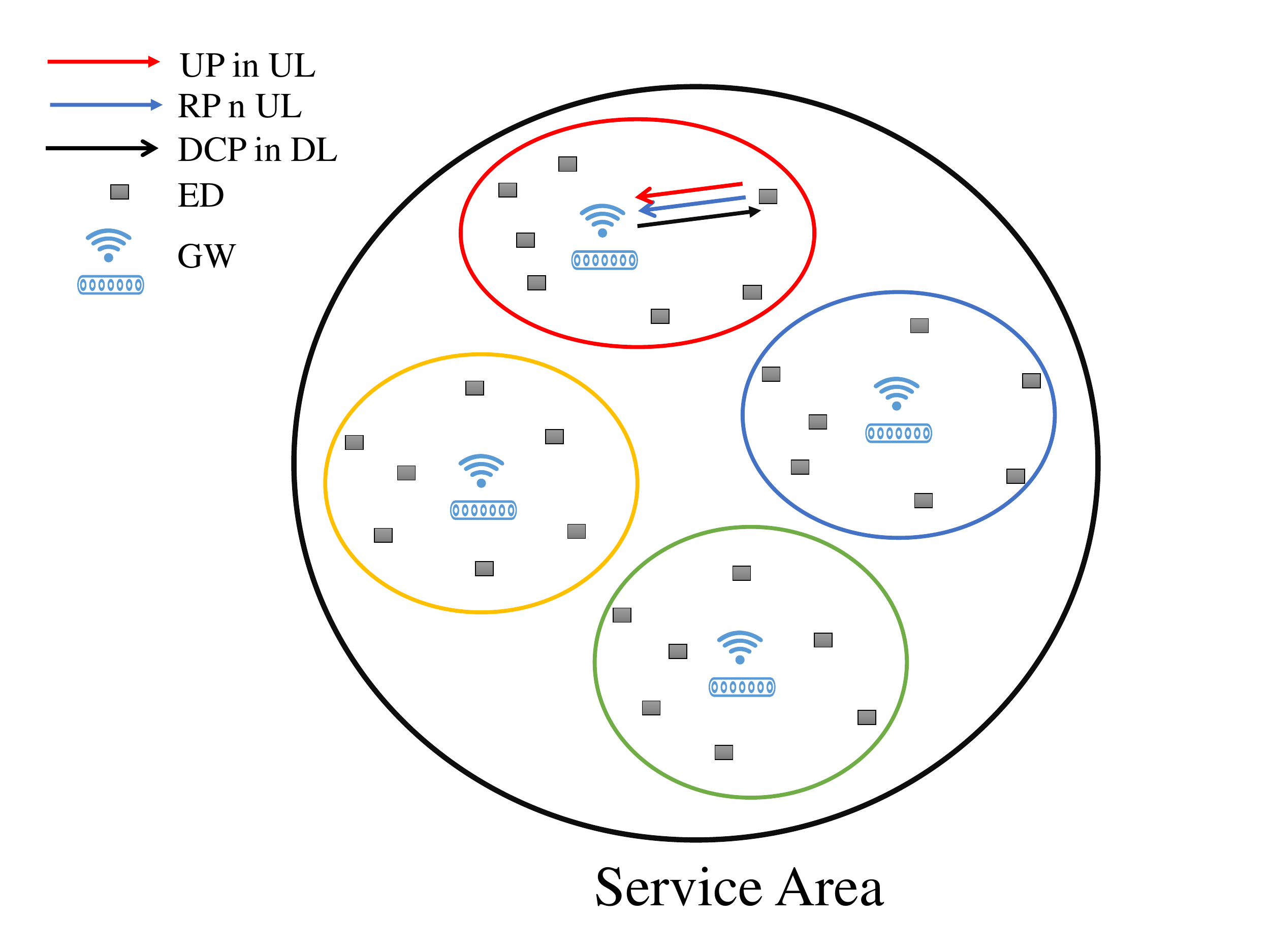}
		\caption{A service area with different clusters represented by different colors.}
		\label{fig:proposed-approach-clusterization}
\end{figure}

The clustering algorithm and the specific localization technology to be adopted in the system are still under investigation and are beyond the scope of this paper. Figure \ref{fig:proposed-approach-clusterization} depicts the overall vision of the system where several GWs are installed inside the service area and the clusters of users are associated to the closest GW (represented by different colors). In the figure DCP stands for downlink control packets, which are regularly transmitted by GWs to the associated EDs as detailed in the next section. Note that ED-GWs association is fully in charge of the server and is only to provide a separated control mechanisms, i.e., it is completely transparent to the EDs which are actually connected to each GW as in standard LoRaWAN.  

\subsection{ Coordination of ED transmissions through Downlink Control Packets (DCPs)}
\label{subsec:DCP description}

We refer to the Class A DL operation in which the Network Server transmits a DL packet to an ED after the reception of an UP packet precisely at the beginning of one of two possible receiving windows. More precisely, the ED opens Class A RX1 and RX2 receiving windows after RECEIVE\_DELAY1 and RECEIVE\_DELAY2 secs respectively. The DL data rate for RX1 depends on the corresponding UL whereas RX2 uses a fixed data rate depending on the region.

In the considered scenario, each node transmits RPs periodically every predetermined (long) time intervals (e.g., several minutes) in UNCONF mode.
We program the server in such a way that for every received RP a corresponding DCP is scheduled. Specifically, the DCP message is intended to control the eventual transmission of UPs. The adopted control mechanism, which is discussed in detail in the next section, acts independently on each cluster since it is highly unlikely that in the considered scenario EDs of different clusters have to transmit an UP at the same time (the potentially dangerous event is local and infrequent). 

Hence, upon the necessity of delivering an UP to the system, the ED transmits according to the control information specified in the last received DCP. More specifically, we opted to choose the UNCONF mode also for UPs. The rationale for this choice will be given in the next section.

\begin{figure}[ht]
	\centering
		 \includegraphics[width=1\columnwidth]{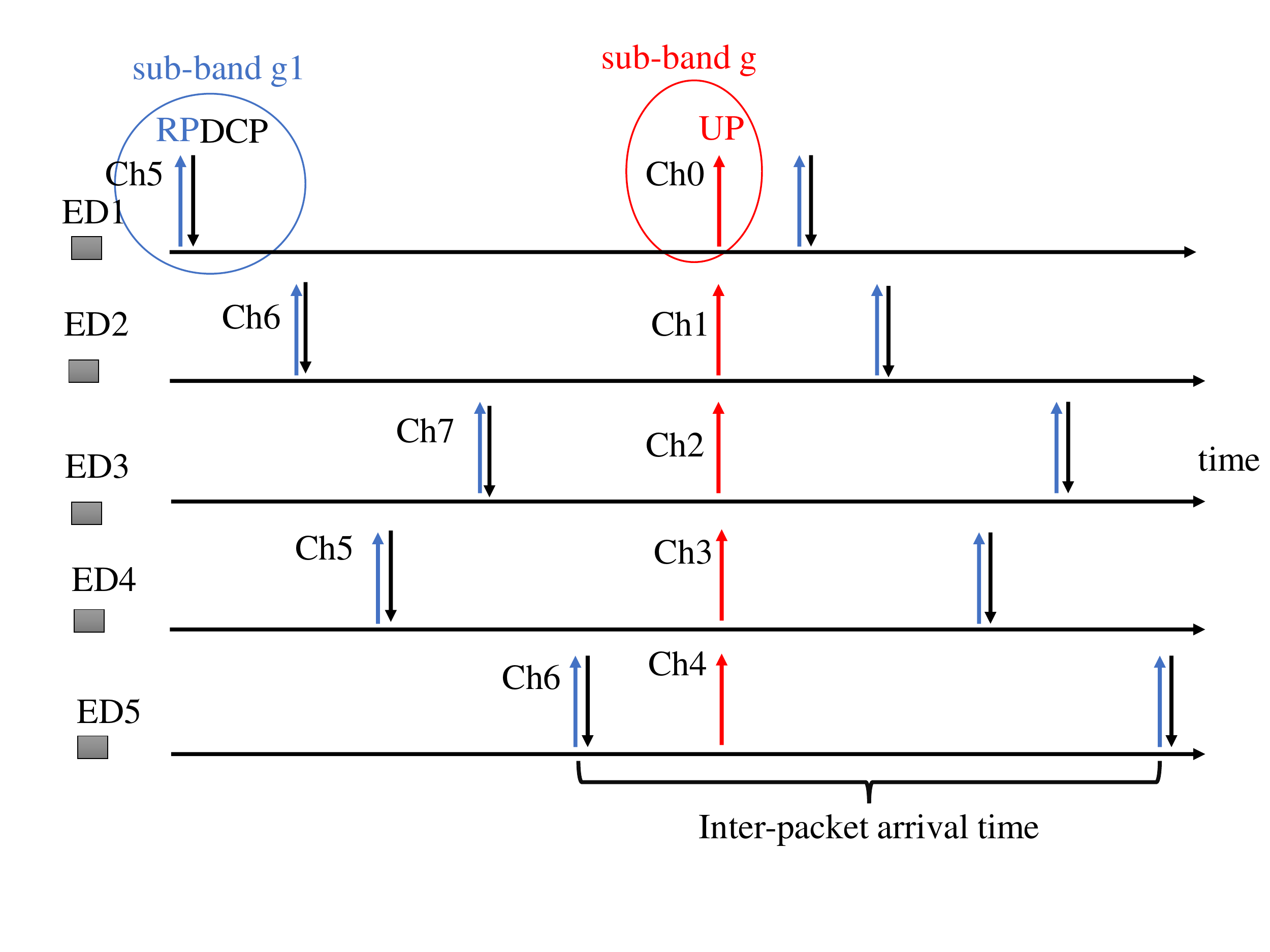}
		\caption{A schematic diagram of transmission of RP, UPs and DCP for 5 EDs belonging to the same cluster where each ED transmits UPs at the same time using different channels from sub-band \textit{g}}
		\label{fig:packet-tx-example}
\end{figure}

One of the important aspects that has to be taken into consideration while designing any LoRaWAN system is to comply with the duty cycle regulations as discussed in Section \ref{SC}. This poses some stringent constrains in the process of allocating the resources to the EDs. Owing to the per sub-band duty cycle regulations, we have various possibilities to assign the resources to the EDs for the next UPs. In particular, one of the feasible choice is to differentiate the sub-bands for the two types of packets, i.e., allocating fixed non-overlapping sub-bands for the RPs and UPs. Indeed, this not only allow the isolation in terms of frequencies but also address the issue of duty cycling in the case when it is necessary to transmit an UP when the time elapsed form the last RP is lower than the minimum time established by duty cycling restrictions. In particular, the server assign different sub-bands for RPs and UPs so that duty cycling restrictions are independently established for the two kinds of transmissions. An illustrative example is given in Figure \ref{fig:packet-tx-example}, where 5 EDs in close proximity, i.e., belonging to the same cluster, are allocated sub-band \textit{g} (Channel (Ch0-4) with frequencies 867.1, 867.3, 867.5, 867.7, 867.9 MHz ) and \textit{g1} (Channel (Ch5-7) with frequencies 868.1, 868.3, 868.5 MHz) for UPs and RPs respectively. To elaborate, each EDs transmit RPs by randomly selecting one of the available frequencies from \textit{g1} whereas the UPs are transmitted using different frequencies in sub-band \textit{g} to avoid collisions. Such frequencies can be selected by each ED according to the last DCP received from the server which is in charge of isolating the UP transmissions of the same cluster. Considering 5 channels in sub-band \textit{g}, it is worth noting that we can allocate a maximum of 5 different channels to 5 EDs in each cluster. However, we also have the possibility to accommodate more users in the cluster by assigning different SFs as shown in the next section.

\subsection{The problem of DL priority}
\label{theoretical_analysis}
   In standard LoRaWAN, the GWs work in half duplex mode only, i.e., they cannot receive and transmit simultaneously. Moreover, in commercial GWs, if there is the need to send a DL message, the reception of any incoming signal is interrupted, i.e., the concurrent UL packet is lost. Accordingly, the mechanism proposed in this paper for coordinating simultaneous UL UPs, which is based on periodic delivery of DCPs, could dramatically affect the PLR of UPs. 
    
    It is then of paramount importance to evaluate the PLR due to GW transmissions. To this aim, it is worth noting that the UL packet is lost by the concurrent DL transmission either when the DL packet is ongoing at the UL packet arrival time, or if it is started during the reception of the UL packet, since the GW gives priority to transmission anyway. Accordingly, denoting by $\tau$ and $D$ the durations of a DCP and of an UP, respectively, the PLR of UPs is equal to the probability that at least one DCP is generated in the interval $\Delta = \tau + D$.  
    To elaborate, in the considered setting DCPs are created as a response of RPs transmitted in the UL by each node. Accordingly, the DCPs arrival process statistics is equivalent to that of RPs generation process. Let then denote by $T$ the rescheduling period set by each ED. Owing to inevitable clock drifts, the actual rescheduling time can be modeled as a random variable (rv) $r = T + \delta$, where $\delta$ is the clock error. 
    
    In the considered scenario we deal with internal clocks which are natively embedded inside the ED microcontroller. This choice allows to save cost, energy, and complexity with respect to external clocks. In this case, it is shown in \cite{Abrardo2020} that the clock errors are unbiased and that they can be reasonably modeled as independent and identically distributed (IID)  Gaussian rvs, i.e., $\delta ~ \text{\textasciitilde}~ \mathcal{N}(0,\sigma)$. Accordingly, also the interarrival times $r$ are IID rvs, i.e., the arrival process of DCPs belong to the class of renewal processes \cite{Parzen1965}. More specifically, we have $r ~ \text{\textasciitilde}~ \mathcal{N}(T,\sigma)$.
    
    From the theory of renewal processes, it is possible to evaluate the time asymptotic density $w(x)$ for the the time elapsed from a generic time till the next arrival, i.e.
    \begin{equation}\label{ITd}
\begin{array}{c}
w(x) = \frac{1}{T}\left(1-F_r(x)\right)
\end{array}
\end{equation}
where $F_r(x)$ is the cdf of $r$. 
In the following, we are interested in evaluating the probability that an UP does not experience any collision with DCPs. To this aim, it is reasonable to assume that different nodes are characterized by independent clocks and independent time delays, and, hence, the probability $P_C$ that the an UP does not experience any collision can be evaluated as the product of individual probabilities of all independent events. To elaborate, let us denote by $N$ the number of EDs and by $\boldsymbol{\tau} = \left\{\tau_1,\tau_2,\ldots,\tau_N\right\}$ the DCP time duration of each node. 
%Hence, we focus on the probability $P_C^{(n)}$ of no collision with node $n$. 
Such terms depends on the SF used by the correspondent nodes to transmit DCPs, i.e., the higher SF the longer $\tau$.  Since the adopted SFs in the UL depend on the channel conditions of each ED, e.g., the distance from the GW, it is reasonable to consider $\boldsymbol{\tau}$ as a set of i.i.d. rvs with individual pdf $f_{{\tau}}(t)$. Similarly, also $D$ depends on the SF adopted by the ED to transmit an UP and, hence, it can be characterized by a given pdf $f_D(y)$.

Accordingly, the probability $P_C$ that the an UP does not experience any collision with DCPs for given $\boldsymbol{\tau} = \left\{\tau_1,\tau_2,\ldots,\tau_N\right\}$ and $D$ is:
\begin{equation}\label{ITd}
\begin{array}{c}
P_C\left(\boldsymbol{\tau},D\right) = \displaystyle\prod\limits_{n=1}^{N}\left(1-\frac{1}{T}\int\limits_{0}^{\tau_n+D}\left(1-F_r(x)\right)dx\right)
\end{array}
\end{equation}
and the marginal probability is:
\begin{equation}\label{ITd}
\begin{array}{c}
P_C =  \left[\displaystyle\int\limits_{{t}}\displaystyle\int\limits_{y}\left(1-\frac{1}{T}\int\limits_{0}^{t+y}\left(1-F_r(x)\right)dx\right)f_{{\tau}}(t) f_D(y) dt dy\right]^N
\end{array}
\end{equation}
with $PLR = 1-P_C$.

In the interesting case where $PLR \ll 1$, the expression in \eqref{ITd} can be manipulated to get an easy to understand approximation of the PLR. To elaborate, when $T \gg \tau+D$ (i.e., small PLR), we have $1-F_r(x) \approx 1$ thus yielding:
\begin{equation}\label{ITda}
\begin{array}{c}
P_C \approx  \left(1-\frac{\mathbb{E}(\tau)+\mathbb{E}(D)}{T}\right)^N\\ 
PLR  \approx N \frac{\mathbb{E}(\tau)+\mathbb{E}(D)}{T}
\end{array}
\end{equation}\label{eq_teo}

\section{Experimental Results and Discussion}
In the following we describe the experimental testbed  to assess the possibility of achieving the required service requirements using the proposed LoRaWAN solution based on DL control and clustering. 
\label{section:results and discussion}
\subsection{Test Scenario}
We consider an indoor testbed at the premises of the Department of Information Engineering and Mathematical Science (DIISM) of the University of Siena. The environment is made of several rooms at the same level and includes the presence of machinery and obstacles, movements of objects and people. In this setting, we deployed several EDs and GWs inside the building for different test scenarios. In particular, we have conducted a preliminary set of tests to evaluate the PLR in the presence of a single ED transmitting in UNCONF mode. In this case, we have verified that the PLR is always well below the required limit of 0.1\% even in the SF7 case. These results are in line with the LoRa coverage expectations and are not reported here for the sake of brevity. Hence, we focus on the results obtained in three different experimental setup characterized by the presence of possible collisions and of concurrent DL transmissions. In all the reported results, we consider only the PLR of UP packets, since RPs are not critical in the considered scenario.   
 We report the parameter settings for the three different scenarios in Table \ref{tab:testbed_setup} where $N_{ED}$, $N_{GW}$, $N_{UP}$, $\Delta t_{RP}$, and $\Delta t_{UP}$ denote the number of EDs, GW, and UPs, and the RPs and UPs inter-packet arrival times, respectively. In order to achieve consistent statistics and reduce the experimental time, we have kept the RPs inter-packet arrival times to relatively low values. Moreover, the UPs are generated by forcing the triggering of gas sensors at regular intervals. 
In the following, we report the description of each scenario, rationale for performing the particular test, the results obtained and the corresponding discussion.

\begin{table}[!htb]
    \caption{Parameter settings for experimental tests}
    \label{tab:testbed_setup}
    \begin{minipage}{.5\linewidth}
      \centering
       \begin{tabular}{|l|c|c|c|c|c|c|c|}
\hline
\textbf{Test} & \textbf{ $N_{ED}$} & \textbf{$N_{GW}$} & \textbf{$N_{UP}$}  & \textbf{ $\Delta t_{RP}$}  & \textbf{ $\Delta t_{UP}$}\\ \hline

\begin{tabular}[c]{@{}l@{}}1   \end{tabular} & 2  & 1 &20000 & 70 sec &Random (120-130 sec) \\ \hline
\begin{tabular}[c]{@{}l@{}}2   \end{tabular} & 8 & 1 &20000 &70 sec & Random (120-130 sec) \\ \hline
\begin{tabular}[c]{@{}l@{}}3   \end{tabular} & 8 & 2 &20000 &70 sec &Random (120-130 sec)\\ \hline

\end{tabular}
    \end{minipage}%
\end{table}

\subsection{Test 1: Analysis of PLR in the presence of collisions}
In the first set of experiments, as reported in Table \ref{tab:testbed_setup}, we deployed two EDs, namely ED1 and ED2 nearly close to each other at distance of approximately 20 m from the GW. Both nodes asynchronously transmit RPs every 70 sec using \textit{Ch5}. In addition, both nodes are triggered to transmit synchronously the UPs at random intervals using \textit{Ch0}. The  SFs adopted for transmitting the UPs are set according to the DCP commands. In this case, we force the users to transmit at the same time using the same channel to assess the possibility of isolating the two transmissions in the SF domain. 

We summarize the results in Tables \ref{tab:plr7-8 },\ref{tab:plr8-9 },\ref{tab:plr9-10 } where we report the PLRs for different cases. We neglect the PLR due to DL transmissions discussed in Section \ref{theoretical_analysis} and which will be separately assessed in Test 2. As expected, the results reveal significant packet losses when transmitting with the same SF. On the other hand, in the case of different SFs, the node transmitting with higher SF has a much lower PLR. i.e., it is able to often capture the packet even in the presence of interference owing to quasi inter-SF  orthogonality. 
Nevertheless, in the SF7-8 case there is a residual probability (slightly higher than the constraint of 0.1\%) that the packet is lost by both EDs, while this probability goes to zero for the SF8-9 and SF9-10 cases. Since it is highly probable that concurrent UPs will report the same information to the server, e.g., a gas concentration is higher than the threshold, in many cases it could be sufficient to receive only one UP to prevent the accident. Under this hypothesis, the 0.1$\%$ constraint can be satisfied when a maximum of 3 nodes in a cluster are assigned the same frequency and different SFs, namely, SFs 8, 9 and 10.

\begin{table}[h]
    \caption{Analysis of PLR between SF7 and SF8}
     \label{tab:plr7-8 }
    \begin{minipage}{.5\linewidth}
      \centering
       \begin{tabular}{|l|c|c|}
\hline
\textbf{Node} & \textbf{SF} & \textbf{PLR (\%)} \\ \hline

\begin{tabular}[c]{@{}l@{}}ED1 \end{tabular} & 7 & 94.3 \\ \hline
\begin{tabular}[c]{@{}l@{}}ED2 \end{tabular} & 7 & 35  \\ \hline
\begin{tabular}[c]{@{}l@{}}Both \end{tabular} & 7 & 29.66  \\ \hline
\end{tabular}
    \end{minipage}%
    \begin{minipage}{.5\linewidth}
      \centering
       \begin{tabular}{|l|c|c|}
\hline
\textbf{Node} & \textbf{SF} & \textbf{PLR (\%)} \\ \hline

%\begin{tabular}[c]{@{}l@{}}Node 1 \end{tabular} & 7 & 5.18 \\
\begin{tabular}[c]{@{}l@{}}ED1 \end{tabular} & 7 & 4.7 \\
\hline
%\begin{tabular}[c]{@{}l@{}}Node 2 \end{tabular} & 8 & 2.87 \\
\begin{tabular}[c]{@{}l@{}}ED2 \end{tabular} & 8 & 3.23 \\
\hline
\begin{tabular}[c]{@{}l@{}}Both \end{tabular} & 7 and 8 & 0.12  \\ \hline
\end{tabular}
    \end{minipage} 

\end{table}

\begin{table}[!htb]
   \caption{Analysis of PLR between SF8 and SF9}
   \label{tab:plr8-9 }
     \begin{minipage}{.5\linewidth}
      \centering
       \begin{tabular}{|l|c|c|}
\hline
\textbf{Node} & \textbf{SF} & \textbf{PLR (\%)} \\ \hline
\begin{tabular}[c]{@{}l@{}}ED1 \end{tabular} & 8 & 17.96 \\ \hline
\begin{tabular}[c]{@{}l@{}}ED2 \end{tabular} & 8 & 84.88 \\ \hline
\begin{tabular}[c]{@{}l@{}}Both \end{tabular} & 8 & 3.46 \\ \hline
\end{tabular}
    \end{minipage}% 
    \begin{minipage}{.5\linewidth}
      \centering
       \begin{tabular}{|l|c|c|}
\hline
\textbf{Node} & \textbf{SF} & \textbf{PLR (\%)} \\ \hline

\begin{tabular}[c]{@{}l@{}}ED1 \end{tabular} & 8 & 32 \\ \hline
\begin{tabular}[c]{@{}l@{}}ED2 \end{tabular} & 9 & 0 \\ \hline
\end{tabular}
    \end{minipage}
\end{table}

\begin{table}[!htb]
   \caption{Analysis of PLR between SF9 and SF10}
   \label{tab:plr9-10 }
    \begin{minipage}{.5\linewidth}
      \centering
       \begin{tabular}{|l|c|c|}
\hline
\textbf{Node} & \textbf{SF} & \textbf{PLR (\%)} \\ \hline

\begin{tabular}[c]{@{}l@{}}ED1 \end{tabular} & 9 & 5.18 \\ \hline
\begin{tabular}[c]{@{}l@{}}ED2 \end{tabular} & 10 & 0 \\ \hline
\end{tabular}
    \end{minipage} 
\end{table}

\subsection{Test 2: analysis of the PLR in the UL due to DL}

We have deployed 8 EDs where ED1-7 transmit asynchronous RPs every $T = 70$ sec (with SF7) whereas ED8 is triggered to transmit also the UP at a random interval of 120-130 sec with SF9. According to the notation used in \eqref{eq_teo} the considered scenario corresponds to $N = 8$, $\mathbb{E}(\tau) = 82.2$ ms, $\mathbb{E}(D) = 267.3$ ms, yielding a predicted PLR of $ 3.9\%$.

In the considered setting we force ED1-7 to use a random channel from sub-band \textit{g1} whereas ED8 is allowed to transmit using \textit{Ch0} from sub-band \textit{g}. Table \ref{tab:PLR in UL due to DL} reports a PLR of 3.66\% which is indeed a very high value and is certainly unacceptable in our case. It is worth noting that this value almost perfectly matches the PLR predicted by \eqref{eq_teo}.

A possible way for overcoming this problem is to duplicate each GW. More specifically, only one of the 2 GWs is configured to transmit the DCPs, so that the other is always free to receive UPs. 

As a matter of fact, there are several papers that propose full-duplex or multi-cast GWs to overcome this problem. However, as discussed in Section \ref{section:related_works}, such feature is not present in off-the-shelf GW solutions.

\begin{table}[!htb]
    \caption{Analysis of PLR in UL due to DL transmission}
    \label{tab:PLR in UL due to DL}
    \begin{minipage}{.5\linewidth}
      \centering
       \begin{tabular}{|l|c|c|}
\hline
\textbf{Node} & \textbf{SF} & \textbf{PLR (\%)} \\ \hline
\begin{tabular}[c]{@{}l@{}}ED8 (UPs)\end{tabular} & 9 & 3.66  \\ \hline
\end{tabular}
    \end{minipage}%
\end{table}

\subsection{Test 3: Analysis of residual loss with two GWs}

\begin{table}[!htb]
    \caption{Analysis of residual loss using double GW for ED8}
    \label{tab:final}
    \begin{minipage}{.5\linewidth}
      \centering
       \begin{tabular}{|l|c|c|}
\hline
\textbf{Test set} & {SF} & \textbf{Number of loss packets} \\ \hline %& \textbf{Sent} \\ \hline

\begin{tabular}[c]{@{}l@{}}1 \end{tabular} & 7 & 10 \\ \hline % &20000 \\ \hline
\begin{tabular}[c]{@{}l@{}}2 \end{tabular} & 8 & 8 \\ \hline % &20000   \\ \hline
\end{tabular}
    \end{minipage}%
\end{table}

 The final test is performed with considering the same scenario of Test 2 in the presence of an additional GW configured to only receive UP packets. Table \ref{tab:final} reports the total number of lost packets for ED8 for two different SFs. It is worth noting that in this case the residual PLR is by far less than 0.1\% even for small SFs (7-8) which confirms that the double GW solution provides the required service levels in the considered scenario.

%%%%%%%%%%%%%%%%%%%%%%%%%%%%%%%%%%%%%%%%%%
\section{Conclusions}
\label{Conclusion}
In this work, we have proposed a LoRaWAN compliant reliable and low-latency solution to fulfill the requirements of a FMAR scenario. To this aim, a low-cost and low-power wearable device is developed to detect the leakage of hazardous and flammable gases. The proposed approach allows to reliably transmit urgent data to the central server. This goal is achieved by leveraging the transmission of DL control messages aimed at avoiding collisions among concurrent transmissions. Finally, we have validated the proposed approach with extensive experimental tests in an industrial-like scenario. Numerical results suggest that LoRaWAN can be exploited to obtain the required level of reliability in the considered scenario. 

%%%%%%%%%%%%%%%%%%%%%%%%%%%%%%%%%%%%%%%%%%
\vspace{6pt} 

%%%%%%%%%%%%%%%%%%%%%%%%%%%%%%%%%%%%%%%%%%
%% optional
%\supplementary{The following are available online at \linksupplementary{s1}, Figure S1: title, Table S1: title, Video S1: title.}


%%%%%%%%%%%%%%%%%%%%%%%%%%%%%%%%%%%%%%%%%%
%\authorcontributions{For research articles with several authors, a short paragraph specifying their individual contributions must be provided. The following statements should be used ``Conceptualization, X.X. and Y.Y.; methodology, X.X.; software, X.X.; validation, X.X., Y.Y. and Z.Z.; formal analysis, X.X.; investigation, X.X.; resources, X.X.; data curation, X.X.; writing---original draft preparation, X.X.; writing---review and editing, X.X.; visualization, X.X.; supervision, X.X.; project administration, X.X.; funding acquisition, Y.Y. All authors have read and agreed to the published version of the manuscript.'', please turn to the  \href{http://img.mdpi.org/data/contributor-role-instruction.pdf}{CRediT taxonomy} for the term explanation. Authorship must be limited to those who have contributed substantially to the work~reported.}
\authorcontributions{``Conceptualization, A.A.; methodology, D.T., A.P., L.P., A.F., and A.A.; software, D.T., L.P.; validation, D.T., A.P., L.P., A.F., and A.A.; formal analysis, A.A.; investigation, D.T.; resources, D.T.; data curation, A.A., A.F.; writing---original draft preparation, D.T.; writing---review and editing, D.T., A.P., L.P. and A.A.; visualization, D.T., A.A.; supervision, A.A.; project administration, A.F. and A.A.; funding acquisition, A.A. All authors have read and agreed to the published version of the manuscript.''}

\acknowledgments{This work has been supported by INAIL, the Italian National Institute for Insurance against Accidents at Work, within the framework of the CP-SEC project: Cyber-Physical system (CPS) for the safety of factories at major accident risk.}

%\conflictsofinterest{Declare conflicts of interest or state ``The authors declare no conflict of interest.'' Authors must identify and declare any personal circumstances or interest that may be perceived as inappropriately influencing the representation or interpretation of reported research results. Any role of the funders in the design of the study; in the collection, analyses or interpretation of data; in the writing of the manuscript, or in the decision to publish the results must be declared in this section. If there is no role, please state ``The funders had no role in the design of the study; in the collection, analyses, or interpretation of data; in the writing of the manuscript, or in the decision to publish the~results''.} 
\conflictsofinterest{The authors declare no conflict of interest.}
%% Optional
%\sampleavailability{Samples of the compounds ... are available from the authors.}

%%%%%%%%%%%%%%%%%%%%%%%%%%%%%%%%%%%%%%%%%%
%% Only for journal Encyclopedia
%\entrylink{The Link to this entry published on the encyclopedia platform.}

%%%%%%%%%%%%%%%%%%%%%%%%%%%%%%%%%%%%%%%%%%
%% Optional
\abbreviations{The following abbreviations are used in this manuscript:\\
\noindent 
\begin{tabular}{@{}ll}
ABP & Activation by Personalization\\
ACK & Acknowledgement\\
ADR & Adaptive Data Rate \\
ADC & Analog to Digital Converter \\
AES & Advanced Encryption Standard\\
API & Application Programming Interface\\
BLE & Bluetooth Low Energy\\
BW &Bandwidth \\
CC & Central Controller \\
Ch & Channel \\
CO & Carbon Monoxide \\
CONF & CONFirmed \\
CR & Coding Rate \\
CSS & Chirp Spread Spectrum \\
DAC & Digital to Analog Converter \\
DCP & Downlink Control Packet \\
DL & Downlink\\
ED & End Device \\
FDMA & Frequency-Division Multiple Access \\
FMAR & Factories at Major Accident Risk \\
GW & Gateway \\
IoT & Internet of Things \\
ISM & Industrial, Scientific and Medical \\
ISPRA & Istituto superiore per la protezione e la ricerca ambientale \\
KPI & Key Performance Indicator \\
LoRa & Long Range \\
LoRaWAN & Long Range Wide Area Network\\
LEL & Lower Explosive Level \\
MAC & Medium Access Control \\
MQTT & Message Queue Telemetry Transport  \\
OTAA & Over-The-Air-Activation \\
O$_2$ & Oxygen \\
PDR & Packet Delivery Rate \\
PLR & Packet Loss Rate \\
QoS & Quality of Service \\
RF & Radio Frequency \\
RP & Regular Packet \\
SD & Secure Digital \\
SF & Spreading Factor \\
TDMA & Time-Division Multiple Access\\
TS & Time Slotted\\
UL & Uplink \\
UNCONF & UNCONFirmed \\
UP & Urgent Packet \\
WiFi & Wireless Fidelity
\end{tabular}}

%%%%%%%%%%%%%%%%%%%%%%%%%%%%%%%%%%%%%%%%%%
%% Optional
%\appendixtitles{no} % Leave argument "no" if all appendix headings stay EMPTY (then no dot is printed after "Appendix A"). If the appendix sections contain a heading then change the argument to "yes".
%\appendixstart
%\appendix
%\section{}
%\subsection{}
%The appendix is an optional section that can contain details and data supplemental to the main text---for example, explanations of experimental details that would disrupt the flow of the main text but nonetheless remain crucial to understanding and reproducing the research shown; figures of replicates for experiments of which representative data are shown in the main text can be added here if brief, or as Supplementary Data. Mathematical proofs of results not central to the paper can be added as an appendix.

%\begin{specialtable}[H] 
%\tablesize{\scriptsize}
%\caption{This is a table caption. Tables should be placed in the main text near to the first time they are~cited.\label{tab1}}
%\tablesize{} % You can specify the fontsize here, e.g., \tablesize{\footnotesize}. If commented out \small will be used.
%\begin{tabular}{ccc}
%\toprule
%\textbf{Title 1}	& \textbf{Title 2}	& \textbf{Title 3}\\
%\midrule
%Entry 1		& Data			& Data\\
%Entry 2		& Data			& Data\\
%\bottomrule
%\end{tabular}
%\end{specialtable}

%\section{}
%All appendix sections must be cited in the main text. In the appendices, Figures, Tables, etc. should be labeled, starting with ``A''---e.g., Figure A1, Figure A2, etc. 

%%%%%%%%%%%%%%%%%%%%%%%%%%%%%%%%%%%%%%%%%%

\reftitle{References}

\end{document}